\documentclass[%
 reprint,
groupedaddress,
 amsmath,amssymb,
 superscriptaddress,
 aps,
 pra,
 longbibliography
]{revtex4-2}

\usepackage{graphicx}
\usepackage{dcolumn}
\usepackage{bm}
\usepackage{braket}
\usepackage{physics}
\usepackage{ulem}
\usepackage{pgfplots}
\pgfplotsset{compat=1.18}
\usepackage{amsmath}	
\usepackage{amsfonts}
\usepackage{braket}
\usepackage{amsthm}
\usepackage{bbm}
\usepackage{comment}
\usepackage{cancel}

\newcommand{\red}[1]{\textcolor{red}{#1}}

\usepackage[unicode=true]{hyperref}
\usepackage{xcolor}

\usepackage[utf8]{inputenc}

\hypersetup{
  colorlinks = true, 
  linkcolor = blue, 
  filecolor = blue, 
  urlcolor = blue, 
  citecolor = blue 
}
\begin{document}

\title{
Local wave-packet spectroscopy of the Drude weight in Tomonaga-Luttinger liquids
}

\author{Takayuki Yokoyama}
\email[]{yokotaka308@hiroshima-u.ac.jp}
\affiliation{Quantum Matter Program, Graduate School of Advanced Science and Engineering, Hiroshima University,
Higashihiroshima, Hiroshima 739-8530, Japan}

\author{Yasuhiro Tada}
\affiliation{Quantum Matter Program, Graduate School of Advanced Science and Engineering, Hiroshima University,
Higashihiroshima, Hiroshima 739-8530, Japan}
\affiliation{Institute for Solid State Physics, University of Tokyo, Kashiwa 277-8581, Japan}


\begin{abstract}
We investigate wave packets generated by a local pulse electric field in a Tomonaga-Luttinger liquid (TLL) with a U(1) charge $q$. Bosonization shows that a local pulse field creates two counterpropagating packets with opposite accumulated charges $\pm qK$ and velocity $v$, where $K$ and $v$ are the TLL parameter and velocity. The packet motion produces a linearly growing charge polarization $P(t)=2 qDt$ in the conducting TLL at non-equilibrium, allowing the Drude weight $D=vK$ to be extracted from real-time local dynamics. We test this local-packet description by time-dependent variational principle simulations of the spin-1/2 XXZ chain. The simulations visualize the density, current, energy, and entanglement profiles in the space-time plane and confirm the predicted packet charge and propagation velocity. We also show that excess energy and entanglement remaining near the pulse region are suppressed in the weak-pulse limit. Finally, the Drude weight obtained from the packet-induced polarization agrees with the exact Bethe-ansatz result, demonstrating local-packet spectroscopy of intrinsic transport in a TLL.
\end{abstract}

\maketitle

\section{Introduction}

The propagation of locally generated charge excitations provides an intuitive real-space picture of transport. This viewpoint is particularly natural in mesoscopic and quantum Hall edge systems, where voltage pulses and quantum point contacts are used to create, partition, and manipulate propagating charge packets~\cite{Levitov1996,KKL2006,Dubois2013,Glattli2017,Feve2007,Kamata2014,Bocquillon2014}. 
In such settings, transport observables are often interpreted in terms of the accumulated effects of individual wave packets. The properties of the packets are usually inferred from repeated pulse measurements of currents, noise, correlations, or interference, rather than by directly following the real-space profile of a single packet.

In contrast, in strongly correlated systems on one-dimensional lattices including quantum magnets, one can ask a more direct question on their many-body properties: what kind of real-space charge or spin packet is generated by a local perturbation, and how do its density, current, energy, and entanglement profiles evolve in time?
In such systems, dynamics is more commonly characterized by correlation functions and dynamical structure factors, as in the standard Tomonaga-Luttinger
liquid (TLL)~\cite{Giamarchi2003}. Transport has also been extensively studied through conductivities and Drude weights~\cite{ShastrySutherland1990, Zotos1999, RMP2021}, including the conductance problem of TLL wires connected to reservoirs~\cite{SafiSchulz1995,Kawabata1996}. 
More recently, hydrodynamic and generalized-hydrodynamic descriptions have provided another powerful framework for one-dimensional transport~\cite{Doyon2020}.
Although these approaches have been highly successful, they do not by themselves specify what kind of real-space packet is created by a local operation. A complementary line of work has therefore studied real-space dynamics after local quantum quenches.
For example, local spin or energy perturbations in XXZ chains have been used to study ballistic and diffusive spreading of spin and energy densities~\cite{Langer2009, Langer2011}, with relevance to thermal transport in one-dimensional quantum magnets~\cite{Hess2007, Sologubenko2007}. 
Related local-quench protocols have been connected to transport coefficients,
including Drude weights, through time-dependent Einstein relations~\cite{Karrasch2014,Karrasch2017},
while wave-packet protocols have been used to analyze nonlinear charge
fractionalization in critical chains~\cite{Ramos2024}.
Wave-packet injection in interacting fermion chains has likewise provided a direct visualization of spin-charge separation and nonlinear fractionalization effects~\cite{Al-Hassanieh2013}. 
These works show that local nonequilibrium dynamics is a useful probe of one-dimensional transport.

The present work addresses a different, more specific question. 
We ask what kind of real-space charge or spin packet is generated by a local U(1) twist, or equivalently by a spatially localized electric-field impulse, and how the resulting packet motion is related to the intrinsic transport coefficient of a TLL. 
Because the perturbation is generated by the conserved density, it is distinct from a generic local density, temperature, or single-particle quench. 
Bosonization~\cite{Giamarchi2003} shows that the accumulated charge of each packet is fixed by the TLL parameter $K$, while the U(1) charge polarization generated by the ballistic separation of the two packets grows with a slope determined by the Drude weight.
We test this local-packet picture using time-dependent variational principle (TDVP)~\cite{Haegeman2011, Haegeman2016} simulations of the spin-\(1/2\) XXZ chain, visualizing the density, current, energy, and entanglement profiles and comparing the Drude weight extracted from the packet-induced polarization with the exact Bethe-ansatz result.

The difference between spatially uniform currents and local wave-packet dynamics is not merely technical but also conceptual. 
Standard transport coefficients, such as the optical conductivity and the Drude weight, are formulated in terms of current correlations or the response to a spatially uniform vector potential within a fixed total U(1) charge sector. 
By contrast, a local electric-field pulse preserves the total charge but creates spatially separated positive and negative U(1) charge accumulations. 
It therefore provides a real-space probe of local charge transport, rather than only a spatially uniform current response. 
The present study shows that these two viewpoints are quantitatively tied together in a gapless TLL: the same stiffness that governs the uniform-current response also controls the dipole motion of locally generated charge packets~\cite{TadaOshikawa2026}. 
This provides a local-packet spectroscopy of the intrinsic Drude weight.

\section{Bosonization description}

We consider a one-dimensional system with a conserved U(1) charge such as the spin-1/2 XXZ chain and interacting fermions under the local pulse electric field or equivalently a U(1) twist.
We provide a bosonization description at the TLL fixed point, where the low-energy Hamiltonian is given by  
\begin{align}
H_{\rm TLL} = \frac{v}{2\pi}\int dx
\left[
K(\partial_x \theta)^2 + \frac{1}{K}(\partial_x \phi)^2
\right].
\end{align}
The parameter $v$ represents the excitation velocity and $K$ is the TLL parameter~\cite{Giamarchi2003}.
The fields $(\phi,\theta)$ are the canonical conjugate pair satisfying $[\phi(x),\partial_x\theta(x')/\pi]=i\delta(x-x')$.

\begin{figure}[t]
\includegraphics[width=\columnwidth]{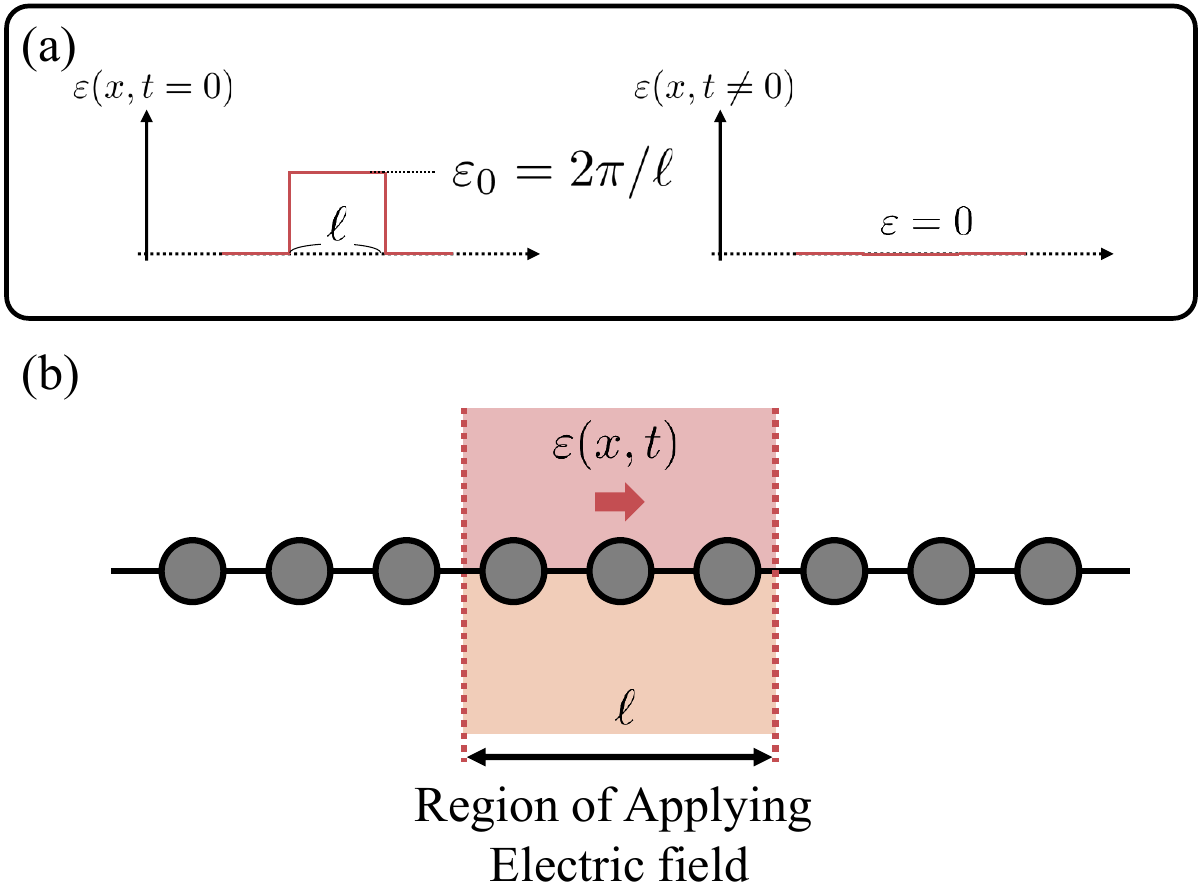}
\caption{Schematic of the local pulse electric field protocol. (a) Spatial profile of the electric field $\mathcal{E} (x,t)$ at $t=0$ and $t \ne 0$, corresponding to the vector potential quench. (b) Illustration of the electric field $\mathcal{E}(x,t)$, applied to the interval $I$ with width $\ell$ in the one-dimensional chain.}
\label{fig:schematic}
\end{figure}

It is useful to distinguish the present protocol from the voltage-pulse protocols commonly discussed in mesoscopic systems~\cite{Levitov1996,KKL2006,Dubois2013,Glattli2017,Feve2007,Kamata2014,Bocquillon2014}. In electron quantum optics and quantum Hall edge setups, a time-dependent voltage applied at a contact or a gate generates a propagating packet whose spatial shape is determined by the temporal profile of the pulse through the propagation of the edge mode. The pulse is therefore local in space but extended in time, and it can be viewed as injecting a controlled packet from an external reservoir or contact.
In the present work, we instead apply a spatially extended but temporally impulsive field, or equivalently a local U(1) twist, directly to the bulk of an isolated lattice TLL (Fig.~\ref{fig:schematic}(a) and (b)). The profile of the generated packet is controlled by the spatial shape of the twist, and the total U(1) charge is conserved. The real-space twist approach has been widely used for studying low-energy spectrum of strongly correlated lattice systems~\cite{LSM1961,AffleckLieb1986,YOA1997,Koma2000,Tasaki2022inbook}, and this method can be used not only for static properties but also for dynamical ones~\cite{TadaOshikawa2026}. In a nonchiral TLL, the local pulse creates a pair of counterpropagating packets with opposite accumulated charges. Thus, our protocol is complementary to the usual temporal-pulse setup: it provides a bulk, real-space probe of local charge transport in a strongly correlated lattice system.
This protocol is also different from generic local quenches of density,
magnetization, or energy profiles~\cite{Langer2009, Langer2011,Karrasch2014,Karrasch2017}.  
Since the perturbation is a gauge twist
generated by the conserved U(1) density, the resulting low-energy packet is
constrained by the TLL parameters, as shown below.

We assume that the low-energy carriers have charge $q$ which is directly coupled to an applied electric field ${\mathcal E}$. 
The charge and current densities are defined as
\begin{align}
\rho(x) = \rho_0 -\frac{q}{\pi}\partial_x \phi(x), \qquad
j_x(x) = \frac{q v K}{\pi}\partial_x \theta(x),
\end{align}
where $\rho_0$ is the average charge density.
In the presence of a vector potential $A$, one has to make the substitution $\partial_x\theta/\pi \to (\partial_x\theta-qA) /\pi$.
This can be implemented by a U(1) twist
\begin{align}
U[\vartheta] = \exp\!\left(i\int dx\, \vartheta(x)n(x)\right)
\label{eq:UTLL}
\end{align}
with $\partial_x\vartheta=qA$ and $n(x) = \rho/q $ is the density-operator, by which the $\theta$-field is shifted as $U\theta(x)U^{\dagger}=\theta(x)-\vartheta(x)$.
The function $\vartheta(x)$ is chosen so that it varies over a region $I$ of width $\ell$ and satisfies $\int_I dx \partial_x \vartheta(x)=2\pi.$
Typically, we consider a simple twist 
\begin{align}
\vartheta(x) = \frac{2\pi }{\ell}(x-x_0)1_{I}(x),
\label{eq:theta}
\end{align}
where $1_I(x)$ is an indicator function, $1_I(x\in I)=1, 1_I(x\notin I)=0$, for the region $I=[x_0,x_0+\ell]$.
The twist function Eq.~\eqref{eq:theta} is often used in the context of the Lieb-Schultz-Mattis theorem and it can generate a low-energy excitation~\cite{LSM1961,AffleckLieb1986,YOA1997,Koma2000,Tasaki2022inbook}.
Note that the present twist operator Eq.~\eqref{eq:UTLL} is spatially local and therefore can generate localized wave packets, which is distinct from the global twist over the whole space.

Now we consider a quench of the vector potential~\cite{Nakagawa2016}, $A(x,t)=A(x)$ for $t\leq 0$ and $A(x,t)=0$ for $t>0$,
for which the electric field is $\mathcal{E}(x,t)=-\partial A(x,t)/\partial t=A_0(x)\delta(t)$ with an amplitude $A_0 (x)=(2\pi/\ell) 1_{I}(x)$ (Fig. ~\ref{fig:schematic}(a) and (b)).
Correspondingly, the current density operator contains $\vartheta(x)$ for $t \leq 0$, while it does not for $t>0$.
Just before the quench, the system is in the ground state and the current density is zero,
\begin{align}
\ev{j_x(x,t=0^-)}
=\frac{q v K}{\pi}\ev{\partial_x(\theta - \vartheta)}=0.
\end{align}
Since the quantum state remains unchanged immediately after the instantaneous quench, we have
\begin{align}
\ev{j_x(x,t=0^+)}
=\frac{q v K}{\pi}\ev{\partial_x\theta}=\frac{q v K}{\pi} \partial_x\vartheta.
\end{align}
In the TLL with right- and left-moving components, dynamics of the $\theta$-field is described by the chiral fields $\phi_R(x-vt)=K\theta(x,t)-\phi(x,t)$ and $\phi_L(x+vt)=K\theta(x,t)+\phi(x,t)$.
Correspondingly, the current density is decomposed into two components as $K\partial_x\theta(x,t)/\pi=\hat{\rho}_R(x-vt)-\hat{\rho}_L(x+vt)$,
where $\hat{\rho}_{R,L}=\pm \partial_x \phi_{R,L} /2\pi$.
Therefore, the current density at time $t>0$ will be given by
\begin{align}
\ev{j_x(x,t)}
=\frac{q v K}{2\pi} \left(\partial_x\vartheta(x-vt) + \partial_x\vartheta(x+vt)\right).
\end{align}
This expression should be valid for the low-energy sector which is described as a TLL, although there may be high energy components beyond the TLL description especially when the amplitude $2\pi/\ell $ is not small.
Indeed, it agrees well with the TDVP results, as discussed in the next section.

Similarly, the expectation value of the excess charge density $\delta \rho=q(\hat{\rho}_R+\hat{\rho}_L)$ is given by
\begin{align}
\ev{\delta \rho(x,t)}
=\frac{q K}{2\pi} \left(\partial_x\vartheta(x-vt) - \partial_x\vartheta(x+vt)\right).
\end{align}
For the electric field profile $\mathcal{E}(x)\sim (2\pi/\ell) 1_I(x)$, this means that the charge packet has a width $W\sim \ell$ and a height $h\sim qK/\ell$.
Correspondingly, the charges of the right- and left-moving packets for time $t \gtrsim \ell /v$ are
\begin{align}
\begin{aligned}
Q_{\rm right} &= \int_{x>x_0+\ell} dx\, \ev{\delta\rho(x,t)} = +qK, \\
Q_{\rm left}  &= \int_{x<x_0} dx\, \ev{\delta\rho(x,t)} = -qK.
\end{aligned}
\label{eq:Q=qk}
\end{align}
The packet charge is controlled by the TLL stiffness through $K$. The packet charge $Q$ is small for a ``hard" system with $K<1$ (repulsive interaction case), while it is large for a ``soft" system with $K>1$ (attractive interaction case).
This is intuitively reasonable and a large charge accumulation will be created for a system with a superconducting (or ferromagnetic) quasi-long-range order.

As a result of the dipole-like charge accumulation $\pm qK$, we have charge polarization 
\begin{align}
P(t)&=\int dx\, x\ev{\delta\rho(x,t)}=vtQ_{\rm right} + (-vt)Q_{\rm left}.
\end{align}
Interestingly, the charge polarization can be rewritten as 
\begin{align}
P(t)=2q{D}t,
\label{eq:Pt_TLL}
\end{align}
where ${D}=vK$ is the Drude weight. 
A similar expression can also be obtained within the standard linear response theory for general lattice models~\cite{TadaOshikawa2026}. Note that, although the present bosonization description is fully consistent with the lattice model result at low-energy, non-Drude terms arising from finite frequency contributions of the conductivity are contained in the latter.
Interestingly, the charge polarization $P(t)$ in Eq.~\eqref{eq:Pt_TLL} is well defined for a conducting system out of equilibrium which is clearly distinguished from the equilibrium charge polarization in an insulator~\cite{KSV1993,VKS1993,Resta2007}.
We stress that the two quantities $P$ and ${D}$ correspond to distinct physical phenomena, since
the charge polarization characterizes a spatially asymmetric charge density profile while the Drude weight is associated with the spatially uniform current.
Conceptually,
the positive and negative charge accumulations in the former are closely related to the energy gap for creating charged excitations.
On the other hand, the charge density is uniform in the latter and thus it is directly related to the energy gap for generating neutral excitations. Thus, the simple relation $P(t)=2q{D}t$ connects the two distinct properties of the system through real-time dynamics due to the local pulse electric field. Since both of the charge-neutral excitations and the charge-carrying excitations are described by a single boson U(1) field in the bosonization theory, the derivation of $P(t)=2q{D}t$ is straightforward as demonstrated above.

\begin{figure*}[tb]
\includegraphics[width=1.0\textwidth]{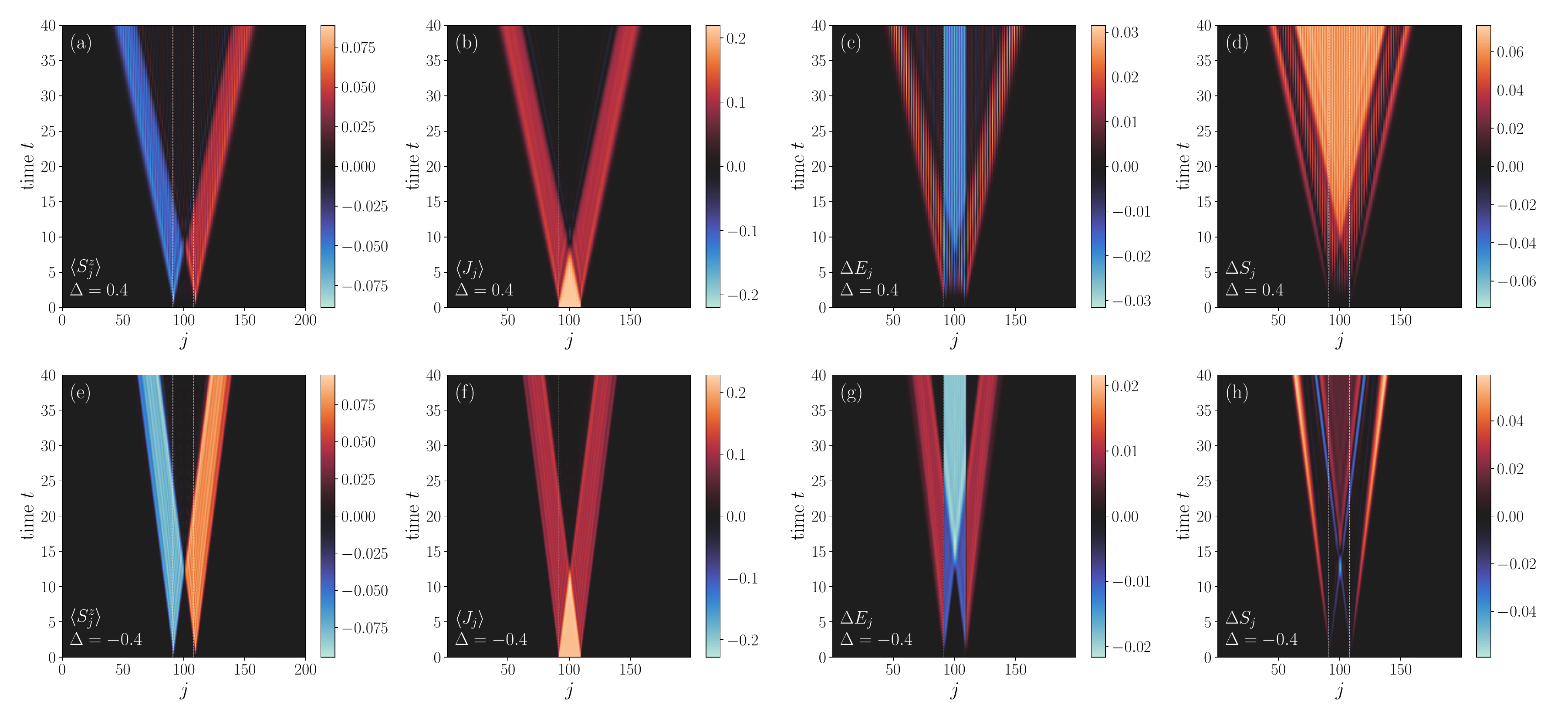}
\caption{Time evolution of (a)(e) spin density, (b)(f) current density, (c)(g) excess energy density, and (d)(h) excess bipartition entanglement entropy in the space-time plane.
The twist width and system size are fixed at $\ell=18$ and $L=200$, respectively. (a)--(d) $\Delta=0.4$ (antiferromagnetic). (e)--(h) $\Delta=-0.4$ (ferromagnetic).}
\label{fig:J04_xt}
\end{figure*}

\section{TDVP calculation}
In this section, 
guided by the bosonization results, we test the following three predictions in the lattice model: (i) the local twist generates two counterpropagating packets, (ii) the accumulated charge is given by \(Q=\pm qK\), and (iii) the packet-induced polarization grows as \(P(t)=2qDt\).
We discuss the $S=1/2$ XXZ spin chain defined on the lattice $\{0,1,\cdots,L-1\}$ under the open boundary condition as a prototypical example of TLLs.
The Hamiltonian is
\begin{align}
H_{\rm XXZ} = J \sum_j \left(
S_j^x S_{j+1}^x + S_j^y S_{j+1}^y
+ \Delta S_j^z S_{j+1}^z
\right),
\label{eq:XXZ}
\end{align}
where the system is gapless for $-1 < \Delta \le 1$.
In this regime, the low-energy physics is described by a TLL with
the exact parameters~\cite{Giamarchi2003,ShastrySutherland1990,Zotos1999}
\begin{align}
v(\Delta) &= \frac{J \pi \sin\gamma}{2\gamma}, \\
K(\Delta) &= \frac{\pi}{2(\pi - \gamma)},
\end{align}
where $\Delta = \cos \gamma$ $(0 \le \gamma < \pi)$.
Since the elementary excitation carries U(1) charge $q=1$, the general results reduce to
\begin{align}
\label{eq:Q}
Q_{\rm right} = +\frac{\pi}{2(\pi - \gamma)}, \qquad
Q_{\rm left}  = -\frac{\pi}{2(\pi - \gamma)}.
\end{align}
As already mentioned, the charge accumulation is large when the system has a ferromagnetic (or superconducting) quasi-long-range order, although the propagating velocity becomes slow.
Note that the XXZ model can be mapped to interacting hard-core bosons and the spin accumulation in the former corresponds to the particle number accumulation in the latter.

\begin{figure*}[t]
\includegraphics[width=1.0\textwidth]{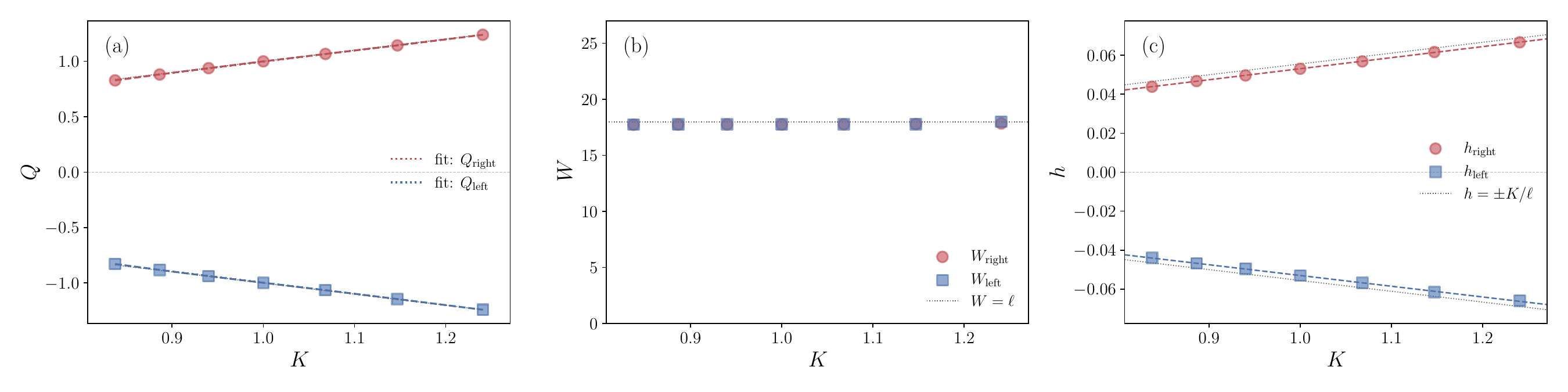}
\caption{
(a) Packet charge $Q_{\rm right}$ (red circles) and $Q_{\rm left}$ (blue squares) as a function of the Luttinger parameter $K$, evaluated from the relaxed states. The dotted lines are linear fits, $Q_{\rm right} = 1.00K$ and $Q_{\rm left} = -1.02K$. (b) Packet width $W$ and (c) packet height $h$ of the right (red circles) and left (blue squares) packets. $W$ is the full width at half maximum of the $\ev{S^z_j}$ after Gaussian smoothing ($\sigma = 2$ sites), and $h$ is the average of the unsmoothed $\ev{S^z_j}$ within this width. Both are averaged over $25 \le t \le 30$. The dashed lines in (c) are linear fits. The gray dotted lines show $W = \ell$ in (b) and $h = \pm K/\ell$ in (c). The twist width is $\ell = 18$ and the system size is $L = 100$.}
\label{fig:Q}
\end{figure*}

The real-time dynamics of the wave packets can be simulated by the TDVP~\cite{Haegeman2011, Haegeman2016}.
We use the open boundary condition for better numerical accuracy and consider a time-scale for which the wave packets do not reach the open ends. We use bond dimensions in the range \(\chi=400\)--\(800\) for which the results are well converged.
In the numerical calculation, we first obtain the ground state $\ket{\psi_0}$ of the Hamiltonian $H_{\rm XXZ}$ by using density-matrix renormalization group (DMRG)~\cite{White1992, DMRG2005} and construct the boosted state $\ket{U}=U\ket{\psi_0}$ by applying the local twist operator 
\begin{align}
U=\exp\left(i\sum_{j\in I}\vartheta_jS^z_j \right),
\label{eq:UXXZ}
\end{align}
where $\vartheta_j=(2\pi/\ell)(j-j_0)1_I(j)$ with $j_0=(L-\ell)/2-1$~ \cite{Yokoyama2026}.
One can show that the excess energy generated by the twist measured from the ground state energy $E^{(0)}$ is $\Delta E=\ev{U|H_{\text{XXZ}}|U}-E^{(0)}=O(1/\ell)$ and it is vanishingly small for large $\ell$.
Besides, the two states $\ket{\psi_0}$ and $\ket{U}$ are orthogonal for large $\ell$ under the periodic boundary condition, and therefore $\ket{U}$ is a low-energy excited state in the gapless phase, which is known as the Lieb-Schultz-Mattis theorem~\cite{LSM1961,AffleckLieb1986,YOA1997,Koma2000,Tasaki2022inbook}.
In the present study, the interval $I=\{j_0,j_0+1,\cdots,j_0+\ell-1\}$ is chosen as the middle region of the system.
Then, we calculate the time evolution of the state $U\ket{\psi_0}$ with the Hamiltonian $H_{\rm XXZ}$ and the physical quantities are evaluated with the time-evolved state 
\begin{align}
\ket{U(t)}=e^{-iH_{\rm XXZ}t}U\ket{\psi_0}.
\label{eq:Ut}
\end{align}
This protocol exactly corresponds to an application of the local pulse electric field $\mathcal{E}_j=(2\pi/\ell) 1_I(j)\delta(t)$ as discussed in the previous section, since the state $U\ket{\psi_0}$ is the ground state of the twisted Hamiltonian $UH_{\rm XXZ}U^{\dagger}$ and the time-evolution is described by the Hamiltonian $H_{\rm XXZ}$ without the twist (vector potential)~\cite{TadaOshikawa2026,Nakagawa2016}.

In Fig.~\ref{fig:J04_xt}, we show a typical example of the calculation results of the spin density $\ev{S_j^z}$, spin current density $\ev{J_j}=iJ/2\ev{S^+_jS^-_{j+1}-S^-_jS^+_{j+1}}$, excess energy density $\Delta E_j=J\ev{S_j^x S_{j+1}^x + S_j^y S_{j+1}^y+ \Delta S_j^z S_{j+1}^z}-E_j^{(0)}$, and excess bipartition entanglement entropy $\Delta S_j=-{\rm tr}_{A_j}\rho_{A_j}\log \rho_{A_j}-S_j^{(0)}$ with $A_j=\{0,1,\cdots,j\}$.
$E_j^{(0)}$ and $S_j^{(0)}$ are the energy density and the entanglement entropy in the ground state, respectively.
There are two wave packets created by the local pulse electric field. The positive spin accumulation propagates to the right and the negative one to the left, which gives a non-zero net current over the entire system.
These qualitative behaviors are common for all the interactions $\Delta$ in the TLL regime, although there are quantitative differences.
Note that the propagation is restricted within the Lieb-Robinson light cone and there is no propagation outside of the light cone~\cite{LR1972,BHV2006,NS2010}.

We also note that the wave packets show no appreciable decay on the simulated time scale in the weak-pulse regime.
This is because the twist Eq.~\eqref{eq:UXXZ} generates low-energy excitations for large $\ell$. 
This behavior is in sharp contrast to wave packets with substantial high-energy components, which can broaden or decay under lattice nonlinearities~\cite{RMP2021,Ramos2024,Al-Hassanieh2013}.

To test the bosonization prediction for the packet charge, we evaluate
$Q_{\rm right}$ and $Q_{\rm left}$ from the density profile once the packets
have relaxed into a steadily propagating form. As shown in
Fig.~\ref{fig:Q}(a), $Q\propto K$ over the whole range studied, with a slope
close to unity ($q=1$), in agreement with
$Q_{\rm right}=-Q_{\rm left}=qK$ [Eq.~\eqref{eq:Q=qk}]. The packet width and
height are $W\simeq\ell$ and $h\simeq K/\ell$ [Fig.~\ref{fig:Q}(b),(c)],
consistent with $\abs{Q} \simeq Wh\simeq K$. Together with the ballistic propagation
in Fig.~\ref{fig:J04_xt}, these results show that the local pulse creates a
controlled low-energy TLL packet rather than a generic density disturbance,
unlike local quenches of density, magnetization, or energy, where
transport is diagnosed from the spreading of the profile
\cite{Langer2009, Langer2011, Karrasch2014, Karrasch2017}.
It would be interesting to extend this diagnostic to exotic TLLs, such as
spin-nematic TLLs~\cite{Hikihara2008} with $q=2$ bound-magnon excitations,
or chiral TLLs at fractional quantum Hall edges~\cite{Wen1990,
GuerreroSuarez2025}, where the packets should carry fractional anyon charge.

\begin{figure}[t]
\includegraphics[width=1.0\columnwidth]{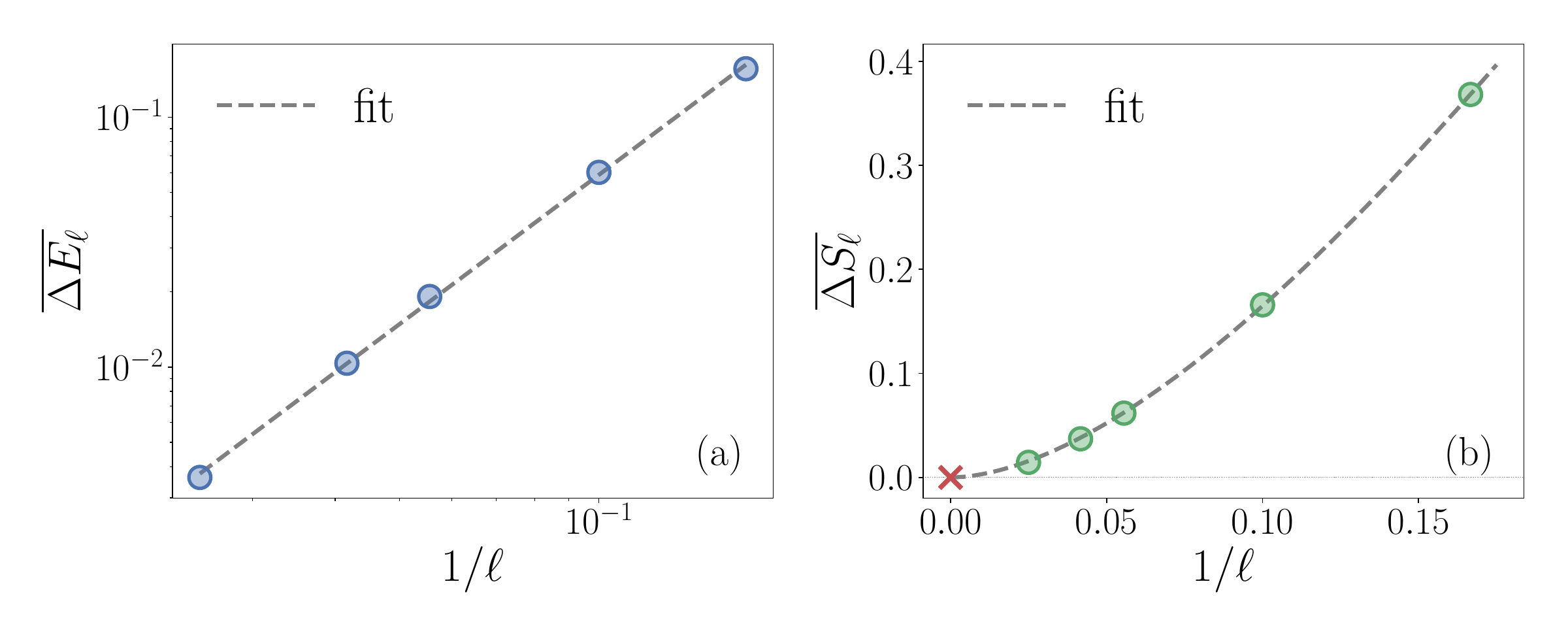}
\caption{
(a) Twist width $\ell$ dependence of the excess energy density.
The fitting line is $\overline{\Delta E_{\ell}} \propto 1/\ell^{1.99}$.
(b) Twist width $\ell$ dependence of the excess entanglement entropy. The fitting is performed using
$\overline{\Delta S_{\ell}} \sim a \log \ell / \ell^2 + b/\ell^2$.
The interaction is $\Delta = 0.4$, and the system size is $L = 100$.
These data are averaged over $30 \le t \le 40 $, by which the wave packets have sufficiently relaxed and their profiles have reached a steady form.
}
\label{fig:J04_l}
\end{figure}

\begin{figure*}[htb]
\includegraphics[width=1.0\linewidth]{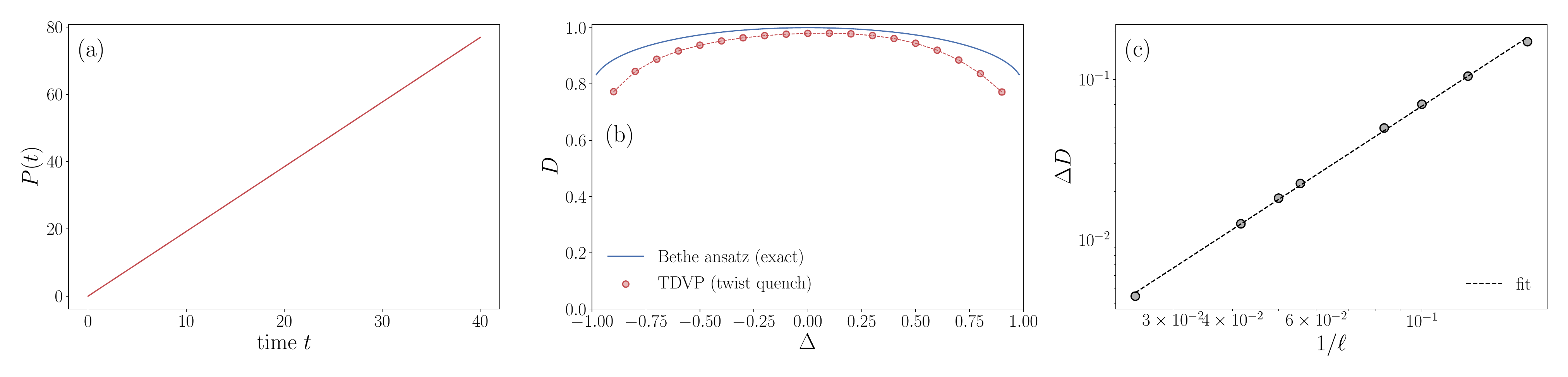}
\caption{(a) Polarization for $\Delta=0.4$ and $\ell=18$.
(b) Drude weight for various $\Delta$ at $\ell=18$, where the solid blue curve is the Bethe-ansatz exact result and the red dots are the TDVP results. (c) Twist width $\ell$ dependence of the difference $\Delta{D}$ for $\Delta=0.4.$
The system size is $L=100$}
\label{fig:DrudeW}
\end{figure*}

Compared with the spin density and current density, there remain some energy density and entanglement entropy in the middle region even in the long time limit.
However, this is due to the finiteness of the twist width $\ell$ or equivalently because the applied electric field $\mathcal{E}\propto 1/\ell$ is not small enough.
To discuss the effects of finite $\ell$, we calculate the average excess energy $\overline{\Delta E}_{\ell}=(1/\ell)\sum_{j\in I}\Delta E_j$ in the twisted region $I$ of width $\ell$. We also calculate the average excess entanglement entropy $\overline{\Delta S}_{\ell}=(1/\ell)\sum_{j \in I}\Delta S_j$, obtained by averaging the excess entanglement entropy $\Delta S_{j}$ of the subsystem $A_j = \{0,\cdots, j | j \in I \}$.
A non-zero $\mathcal{E}$ may give a local energy proportional to $\mathcal{E}^2\sim 1/\ell^2$ and indeed,
as shown in Fig.~\ref{fig:J04_l} (a) and (b), the excess energy behaves as $\overline{\Delta E}_{\ell}\sim 1/\ell^2 \to 0$ for $\ell \to \infty$.
Similarly, the excess entanglement entropy behaves as $\overline{\Delta S_{\ell}}\sim a\log\ell/\ell^2+b/\ell^2\to0$ for $\ell\to\infty$, where $a,b$ are non-universal constants.
Since each wave packet is contained in the region $A_j$ ($j \in I$) or its complement, the vanishing excess entanglement entropy means that the two wave packets are not entangled in the large $\ell$ limit.
The residual excess entanglement entropy ($\overline{\Delta S}_{{\ell}}$) between the wave packets for finite $\ell$ may be regarded as a signature of deviation from the TLL, since no entanglement will be generated between the left- and right-moving components within a pure TLL. Indeed, the exactly vanishing entanglement in a pure TLL can easily be understood from the twist operator in the Heisenberg representation,
\begin{align}
U_{\text{H}}(t) 
&= e^{iH_{\rm TLL}t}Ue^{-iH_{\rm TLL}t} \nonumber\\
&= e^{i\int dx\, \vartheta(x)\hat{\rho}_L(x+vt)} e^{i\int dx\, \vartheta(x)\hat{\rho}_R(x-vt)}.
\end{align}
Therefore, the finite excess entanglement observed in the lattice calculation should originate from components of the boosted state outside the ideal TLL fixed-point description.
When $\ket{U(t)}$ (Eq.~\eqref{eq:Ut}) is written as a superposition of a TLL state $\ket{\psi_{\rm TLL}(t)}$ and a non-TLL state $\ket{\psi_{\rm nonTLL}(t)}$,
\begin{align}
\ket{U(t)}=a\ket{\psi_{\rm TLL}(t)} + b\ket{\psi_{\rm nonTLL}(t)},
\end{align}
the coefficient $b$ is expected to be $b\sim \mathcal{E}\sim 1/\ell$ in the first order perturbation.
Then, it is natural to consider that the change of a Schmidt value $\lambda$ is governed by the weight of the non-TLL state, $\Delta (\lambda_{n})^2 \sim |b|^2\sim 1/\ell^2$.
For Schmidt values that remain finite in the weak-pulse limit, the entropy correction is analytic and scales as $1/\ell^2$. The logarithmic factor originates from the singular behavior of $-\lambda ^2 \log\lambda^2$ near $\lambda=0$. Since a gapless TLL has a dense low-lying entanglement spectrum, finite-pulse corrections can populate Schmidt sectors with $\lambda^2 \lesssim 1/\ell^2$, producing a contribution of order $\log\ell/\ell^2$.

The entanglement dynamics found here is distinct from the entanglement generation in standard global or local quenches~\cite{CalabreseCardy2005,CalabreseCardy2007,EislerPeschel2007,Lauchli2008,Gruber2021}. In those problems, entanglement growth is commonly interpreted as the emission and propagation of entangled quasiparticles. By contrast, our local pulse is a local U(1) twist generated by the conserved density. Within the ideal TLL theory, its time evolution factorizes into independent left- and right-moving unitary operators, so that the separated packets do not generate additional entanglement across the center bipartition. The finite excess entanglement observed in the TDVP data is therefore a finite-$\ell$ correction, or equivalently a signature of deviations from the ideal TLL fixed point, rather than an intrinsic entanglement production mechanism of the TLL packet itself.


\begin{figure*}
\includegraphics[width=\textwidth]{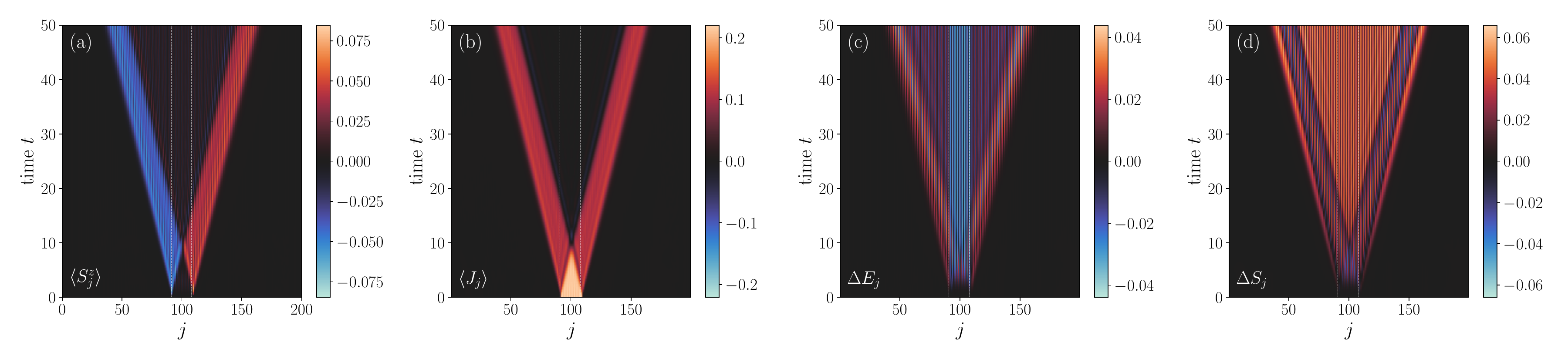}
\caption{Time evolution of (a) spin density, (b) current
density, (c) excess energy density, and (d) excess bipartition
entanglement entropy in the space-time plane for the
non-integrable $J_1$-$J_2$ XXZ chain, with next-nearest-neighbor
coupling $J_2 = 0.1 J$, interaction $\Delta = 0.4$, twist width $\ell = 18$, and system size $L=200$.}
\label{fig:J2_01_J04_xt}
\end{figure*}


We now turn to one of the central consequences of the local-packet picture, namely the relation between the packet motion and the Drude weight. For this purpose, we consider the charge polarization generated by the local pulse. We emphasize again that the polarization discussed here is not the ground-state polarization of an equilibrium insulator. In the Berry-phase theory of polarization, the polarization is defined for an insulating ground state and is related to the charge transported by an adiabatic and spatially uniform current~\cite{KSV1993,VKS1993,Resta2007}. In contrast, the present quantity is a nonequilibrium polarization of a conducting system. It is defined at each time from the real-space density profile of the dynamically propagating packets,
\begin{align}
P(t) = \sum_j j \left(\ev{U(t)|S^z_j|U(t)} - \ev{\psi_0|S^z_j|\psi_0}\right).
\end{align}
Thus, \(P(t)\) measures the instantaneous dipole moment produced by the spatial separation of the positive and negative charge accumulations.
From the packet viewpoint, the physical meaning of this quantity is transparent. The local pulse creates a dipole-like configuration: the right-moving packet carries charge \(+K\), while the left-moving packet carries charge \(-K\), where the U(1) charge is $q=1$ in the XXZ model. Since the two packets propagate ballistically with velocity \(v\), the dipole moment grows linearly in time. The nontrivial point is that the coefficient of this growth is not an independent packet parameter, but is precisely the Drude weight, $P(t)=2Dt$ [Eq.\eqref{eq:Pt_TLL}] with $D=vK$ as predicted by the bosonization argument. 
The Drude weight is usually introduced as a response coefficient to a uniform flux or a spatially uniform electric field. It is therefore not obvious a priori that it should appear in a polarization defined from an inhomogeneous nonequilibrium density profile. The above relation shows that, at the TLL fixed point, the same bulk stiffness that governs the uniform-current response also governs the ballistic separation of locally generated charge packets.
We should distinguish this relation from previous local-quench approaches to transport coefficients. 
In particular, finite-temperature local quenches have been related to Drude weights through time-dependent Einstein relations, where the transport coefficient is extracted from the growth of spatial variances of density profiles~\cite{Karrasch2014,Karrasch2017}. Our relation has a different structure.
At zero temperature and at the TLL fixed point, the local U(1) twist creates two chiral packets with charges fixed by $K$. 
The Drude weight then appears directly as the coefficient of the dipole motion,
$P(t)=2qDt$ [Eq.~\eqref{eq:Pt_TLL}], rather than through the second moment of a spreading profile. 
Thus, the protocol probes the same intrinsic stiffness as the uniform-current response, but through a first-moment, polarization-based dynamics of a controlled pair of TLL packets.

We show an example of the calculated polarization in Fig.~\ref{fig:DrudeW} (a) and find that $P(t)\propto t$ as discussed within the bosonization description in the previous section.
The coefficient  $P(t)/2t$ is given by the Drude weight [Eq.~\eqref{eq:Pt_TLL}] according to the bosonization theory. 
As seen in Fig.~\ref{fig:DrudeW} (b), they are in good agreement over the whole region of $\Delta$, and difference between the calculated result $D_{\rm TDVP}$ and the exact Bethe-ansatz result $D_{\rm exact}=J\pi^2 \sin\gamma/ [4\gamma (\pi-\gamma)]$ with $\Delta=\cos\gamma$~\cite{Zotos1999} is a few percent for the twist width $\ell=18$. 
The difference given by $\Delta D = \abs{D_{\rm TDVP} - D_{\rm exact}}$ is suppressed for large $\ell$ and it approaches zero as shown in Fig.~\ref{fig:DrudeW} (c).
The agreement over the TLL regime, together with the suppression of the deviation for larger twist width, demonstrates that the local-packet protocol provides a real-space nonequilibrium spectroscopy of the intrinsic Drude weight.

Finally, we perform similar calculations also for the $J_1$-$J_2$ XXZ model, where the next-nearest neighbor interaction term $J_2\sum_j(S^x_jS^x_{j+2}+S^y_jS^y_{j+2}+\Delta S^z_jS^z_{j+2})$ is added to the standard XXZ model Eq.~\eqref{eq:XXZ}.
This model is not a solvable model and can be used to confirm that the observed behaviors in the solvable XXZ model are not specific to the integrability.
We find that the wave-packet picture remains valid even in the non-integrable  spin-$1/2$ $J_1$-$J_2$ XXZ model and the bosonization description is applicable (see Appendix for details). Therefore, generation of the ballistic wave packets by the local pulse field is a generic property of TLLs.

\section{summary}

In summary, we have studied real-time wave-packet dynamics induced by a local pulse electric field in TLLs. Within the bosonization description, a local pulse generates two counterpropagating packets with opposite accumulated U(1) charges. The packet velocity is given by the TLL velocity $v$, while the accumulated charge $Q$ is controlled by the TLL parameter $K$. As a result, the induced polarization grows linearly in time as $P(t)=2qDt$, where $ D=vK$ is the Drude weight.

We have tested this picture by TDVP simulations of the spin-1/2 XXZ chain. The numerical results visualize the propagation of the density and current packets in the space-time plane and confirm that the packet width, height, and accumulated charge are consistent with the bosonization prediction. We have also examined the energy density and bipartition entanglement entropy generated by the local pulse. The residual energy density and excess entanglement in the twist region decrease with increasing twist width, indicating that the wave-packets approach the ideal TLL behavior in the weak-pulse limit.

Finally, we have extracted the Drude weight from the packet-induced polarization. The value obtained from the TDVP result $P(t)/2qt$ with $q=1$ agrees well with the exact Bethe-ansatz result for the XXZ chain, and the deviation is suppressed as the twist width increases. This agreement shows that the local-packet protocol provides a real-space way to probe the intrinsic Drude weight of a lattice TLL. Thus, the local pulse response gives a complementary perspective on transport: it visualizes charge transport as the ballistic separation of local charge accumulations, while quantitatively connecting this motion to the conventional uniform-current response.
The present results demonstrate that the Drude weight is not only a coefficient of a uniform current response but also governs the real-time motion of locally generated charge packets in a TLL.

\begin{acknowledgments}
We thank Masaki Oshikawa, Ryotaro Sano, and Hal Tasaki for fruitful discussions. The DMRG and TDVP calculations were performed with the use of TeNPy library~\cite{Hauschild2024} (version 1.1.0). The computation in this work has been partly done using the facilities of the Supercomputer Center, the Institute for Solid State Physics, the University of Tokyo. This study is supported by JST SPRING Grant No JPMJSP2132 and JSPS KAKENHI Grant No. 26K06946.
\end{acknowledgments}

\appendix

\begin{figure}[t]
  \centering
  \includegraphics[width=1.0\columnwidth]{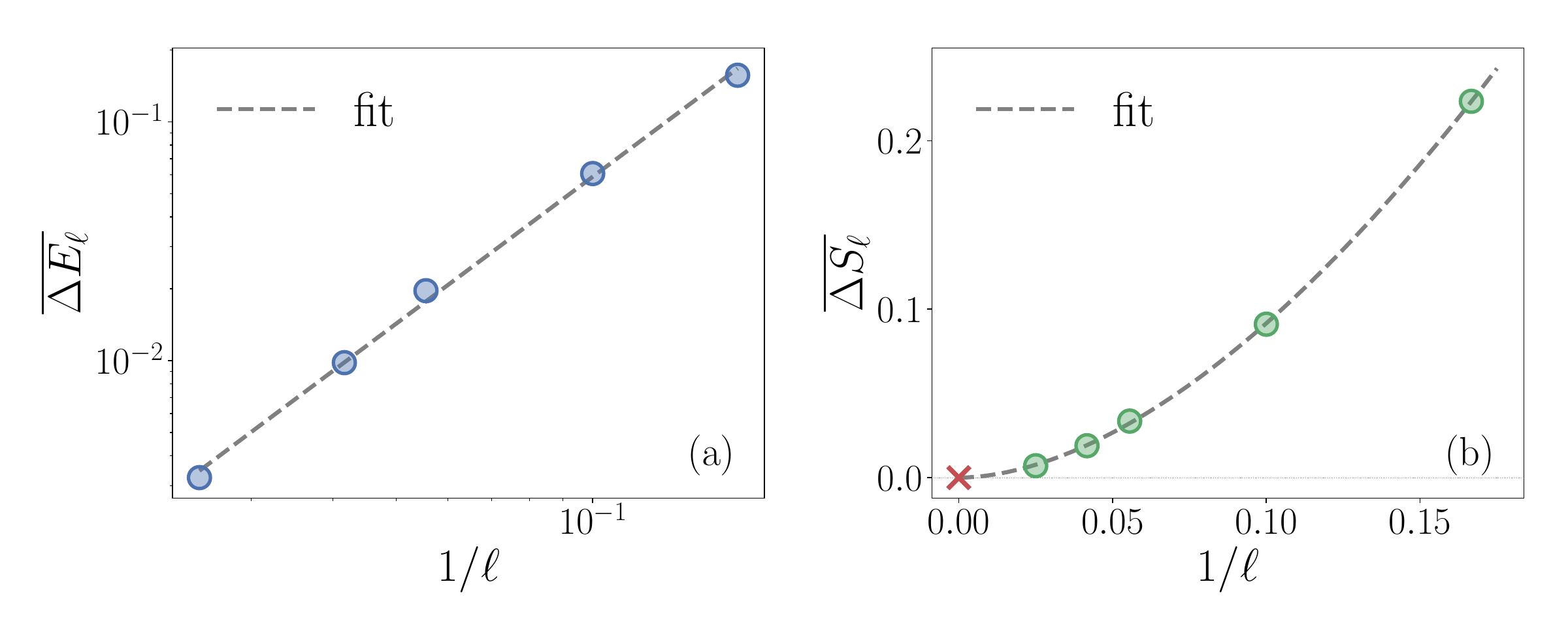}
    \caption{%
    (a)~Twist width $\ell$ dependence of the excess energy density. 
    The fitting line is $\overline{\Delta E_\ell} \propto 1/\ell^{2.05}$.
    (b)~Twist width $\ell$ dependence of the excess entanglement entropy.
    The fitting is performed by
    $\overline{\Delta S_{\ell}} \sim a\,\ell^{-2}\log\ell + b\,\ell^{-2}$.
    The interaction is $J_2 = 0.1J, \Delta=0.4$ and the system size is $L=100$.
    These data are averaged over $30 \le t \le 50$, by which the wave packets have
    sufficiently relaxed and their profiles have reached a steady form.
  }
  \label{fig:J2_01_l}
\end{figure}

\section{$J_1$-$J_2$ model}
In this section, to verify that the local-packet picture is not specific to integrable models, we repeat the same protocol for the non-integrable  spin-$1/2$ $J_1$-$J_2$ XXZ chain, in which the next-nearest-neighbor interaction
\begin{equation}
H^{J_2}_{\text{XXZ}} = J_2 \sum_j \left( S_j^x S_{j+2}^x + S_j^y S_{j+2}^y
+ \Delta S_j^z S_{j+2}^z \right)
\end{equation}
is added to the nearest-neighbor Hamiltonian $H_{\rm XXZ}$ [Eq.~\eqref{eq:XXZ}], with the same anisotropy $\Delta$. 
We prepare the ground state of $H_{\rm XXZ} + H^{J_2}_{\text{XXZ}}$, apply the same local $U(1)$ twist Eq.~\eqref{eq:UXXZ} with twist width $\ell$, and evolve the boosted state $\ket{U(t)}$ with TDVP.

\begin{figure}[t]
\includegraphics[width=1.0\columnwidth]{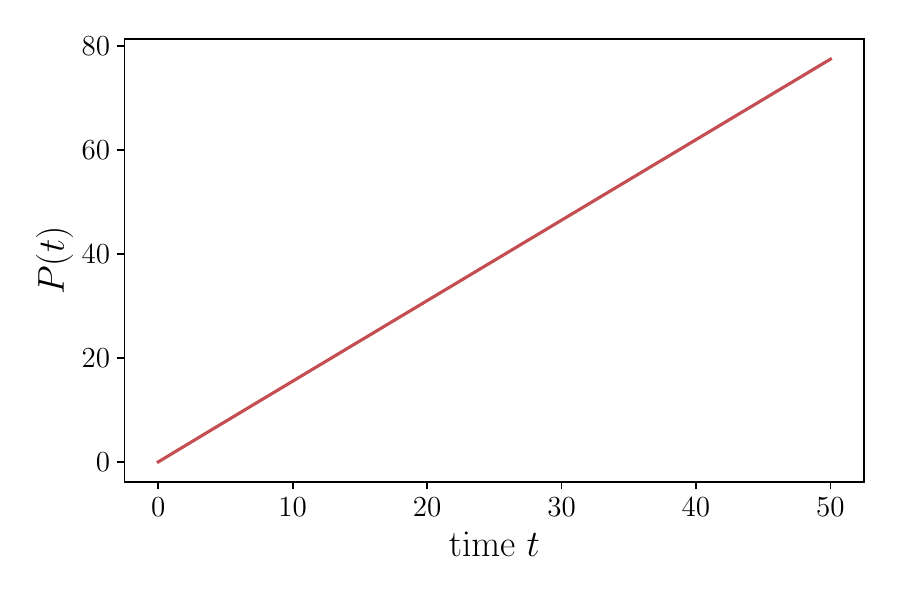}
\caption{Polarization $P(t)$ as a function of time $t$ for
the anisotropy $\Delta = 0.4$, the next-nearest-neighbor interaction $J_2=0.1J$, the system size $L=200$, and twist width $\ell = 18$.}
\label{fig:J2_01_P}
\end{figure}

Figure ~\ref{fig:J2_01_J04_xt} shows the resulting space-time plots of the spin-density $\langle S_j^z\rangle$, current density$\langle J_j\rangle$, excess energy density $\Delta E_j$, and excess entanglement entropy $\Delta S_j$ to be compared directly with Fig.~\ref{fig:J04_xt} for the pure integrable XXZ chain ($J_2=0$). 
The local twist creates two ballistically propagating wave packets. We also find that the residual excess energy $\overline{\Delta E}_{\ell}$ and excess entanglement entropy $\overline{\Delta S}_{\ell}$ in the twisted region decay with increasing $\ell$ in the same way as in the integrable XXZ chain (Fig.~\ref{fig:J2_01_l}(a) and (b)). These results suggest that the bosonization argument holds not only for integrable systems but also for non-integrable ones. The generation of ballistic wave packets and their transport by a local pulse field are generic properties of Tomonaga-Luttinger liquids, not specific to an integrable system.

We also compute the polarization $P(t)$ (Eq.~\eqref{eq:Pt_TLL}) for the $J_1$-$J_2$ chain and confirm that it grows linearly in time, $P(t) \propto t$ (Fig.~\ref{fig:J2_01_P}). 
Because the $J_1$-$J_2$ XXZ chain is not integrable, we do not have an independent exact value of $D$ for comparison.  
Nevertheless, we can still evaluate the Drude weight from the linear growth of the polarization, which is identified with the
Drude weight entering the uniform current response according to the TLL argument in the main text.

\bibliography{main}

@misc{TadaOshikawa2026,
note = {Yasuhiro Tada and Masaki Oshikawa, to be submitted.}
}

@book{Giamarchi2003,
author = {Thierry Giamarchi},
title = {Quantum Physics in One Dimension},
publisher = {Oxford University Press},
year = {2003}
}

@article{Feve2007,
author = {G. Fève  and A. Mahé  and J.-M. Berroir  and T. Kontos  and B. Plaçais  and D. C. Glattli  and A. Cavanna  and B. Etienne  and Y. Jin },
title = {An On-Demand Coherent Single-Electron Source},
journal = {Science},
volume = {316},
number = {5828},
pages = {1169-1172},
year = {2007},
doi = {10.1126/science.1141243},
URL = {https://www.science.org/doi/abs/10.1126/science.1141243}
}

@article{Levitov1996,
    author = {Levitov, Leonid S. and Lee, Hyunwoo and Lesovik, Gordey B.},
    title = {Electron counting statistics and coherent states of electric current},
    journal = {Journal of Mathematical Physics},
    volume = {37},
    number = {10},
    pages = {4845-4866},
    year = {1996},
    month = {10},
    issn = {0022-2488},
    doi = {10.1063/1.531672},
    url = {https://doi.org/10.1063/1.531672}
}

@article{KKL2006,
  title = {Minimal Excitation States of Electrons in One-Dimensional Wires},
  author = {Keeling, J. and Klich, I. and Levitov, L. S.},
  journal = {Phys. Rev. Lett.},
  volume = {97},
  issue = {11},
  pages = {116403},
  numpages = {4},
  year = {2006},
  month = {Sep},
  publisher = {American Physical Society},
  doi = {10.1103/PhysRevLett.97.116403},
  url = {https://link.aps.org/doi/10.1103/PhysRevLett.97.116403}
}

@article{Dubois2013,
  title   = {Minimal-excitation states for electron quantum optics using levitons},
  author  = {Dubois, Jonathan and Jullien, Thibaut and Portier, Fr{\'e}d{\'e}ric and Roche, Patrick and Cavanna, Andrea and Jin, Yong and Wegscheider, Werner and Roulleau, Preden and Glattli, D. Christian},
  journal = {Nature},
  volume  = {502},
  pages   = {659--663},
  year    = {2013},
  doi     = {10.1038/nature12713}
}

@article{Glattli2017,
author = {Glattli, D. Christian and Roulleau, Preden S.},
title = {Levitons for electron quantum optics},
journal = {physica status solidi (b)},
volume = {254},
number = {3},
pages = {1600650},
doi = {https://doi.org/10.1002/pssb.201600650},
url = {https://onlinelibrary.wiley.com/doi/abs/10.1002/pssb.201600650},
year={2017}
}

@article{Kamata2014,
  title   = {Fractionalized wave packets from an artificial Tomonaga--Luttinger liquid},
  author  = {Kamata, Hiroshi and Kumada, Norio and Hashisaka, Masayuki and Muraki, Koji and Fujisawa, Toshimasa},
  journal = {Nature Nanotechnology},
  volume  = {9},
  pages   = {177--181},
  year    = {2014},
  doi     = {10.1038/nnano.2013.312}
}

@article{Bocquillon2014,
author = {Bocquillon, Erwann and Freulon, Vincent and Parmentier, François D. and Berroir, Jean-Marc and Plaçais, Bernard and Wahl, Claire and Rech, Jérome and Jonckheere, Thibaut and Martin, Thierry and Grenier, Charles and Ferraro, Dario and Degiovanni, Pascal and Fève, Gwendal},
title = {Electron quantum optics in ballistic chiral conductors},
journal = {Annalen der Physik},
volume = {526},
number = {1-2},
pages = {1-30},
doi = {https://doi.org/10.1002/andp.201300181},
url = {https://onlinelibrary.wiley.com/doi/abs/10.1002/andp.201300181},
year = {2014}
}

@article{ShastrySutherland1990,
  title = {Twisted boundary conditions and effective mass in Heisenberg-Ising and Hubbard rings},
  author = {Shastry, B. Sriram and Sutherland, Bill},
  journal = {Phys. Rev. Lett.},
  volume = {65},
  issue = {2},
  pages = {243--246},
  numpages = {0},
  year = {1990},
  month = {Jul},
  publisher = {American Physical Society},
  doi = {10.1103/PhysRevLett.65.243},
  url = {https://link.aps.org/doi/10.1103/PhysRevLett.65.243}
}

@article{Zotos1999,
  title = {Finite Temperature Drude Weight of the One-Dimensional Spin- $1/2$ Heisenberg Model},
  author = {Zotos, X.},
  journal = {Phys. Rev. Lett.},
  volume = {82},
  issue = {8},
  pages = {1764--1767},
  numpages = {0},
  year = {1999},
  month = {Feb},
  publisher = {American Physical Society},
  doi = {10.1103/PhysRevLett.82.1764},
  url = {https://link.aps.org/doi/10.1103/PhysRevLett.82.1764}
}

@article{SafiSchulz1995,
  title = {Transport in an inhomogeneous interacting one-dimensional system},
  author = {Safi, I. and Schulz, H. J.},
  journal = {Phys. Rev. B},
  volume = {52},
  issue = {24},
  pages = {R17040(R)--R17043(R)},
  numpages = {0},
  year = {1995},
  month = {Dec},
  publisher = {American Physical Society},
  doi = {10.1103/PhysRevB.52.R17040},
  url = {https://link.aps.org/doi/10.1103/PhysRevB.52.R17040}
}

@article{Kawabata1996,
author = {Kawabata ,Arisato},
title = {On the Renormalization of Conductance in  Tomonaga-Luttinger Liquid},
journal = {Journal of the Physical Society of Japan},
volume = {65},
number = {1},
pages = {30-32},
year = {1996},
doi = {10.1143/JPSJ.65.30}
}

@article{RMP2021,
  title = {Finite-temperature transport in one-dimensional quantum lattice models},
  author = {Bertini, B. and Heidrich-Meisner, F. and Karrasch, C. and Prosen, T. and Steinigeweg, R. and \ifmmode \check{Z}\else \v{Z}\fi{}nidari\ifmmode \check{c}\else \v{c}\fi{}, M.},
  journal = {Rev. Mod. Phys.},
  volume = {93},
  issue = {2},
  pages = {025003},
  numpages = {71},
  year = {2021},
  month = {May},
  publisher = {American Physical Society},
  doi = {10.1103/RevModPhys.93.025003},
  url = {https://link.aps.org/doi/10.1103/RevModPhys.93.025003}
}

@article{Doyon2020,
	title = {Lecture notes on Generalised Hydrodynamics},
	pages = {18},
	author = {Doyon, Benjamin},
	journal = {SciPost Phys. Lect. Notes},
	year = {2020},
	publisher = {SciPost},
	doi = {10.21468/SciPostPhysLectNotes.18},
	url = {https://scipost.org/10.21468/SciPostPhysLectNotes.18}
}

@article{LSM1961,
title = {Two soluble models of an antiferromagnetic chain},
journal = {Ann. Phys.},
volume = {16},
number = {3},
pages = {407-466},
year = {1961},
issn = {0003-4916},
doi = {https://doi.org/10.1016/0003-4916(61)90115-4},
url = {https://www.sciencedirect.com/science/article/pii/0003491661901154},
author = {Elliott Lieb and Theodore Schultz and Daniel Mattis}
}

@article{AffleckLieb1986,
title = {A proof of part of {H}aldane's conjecture on spin chains},
journal = {Lett. Math. Phys.},
volume = {12},
pages = {57--69},
year = {1986},
url = {https://doi.org/10.1007/BF00400304},
author = {Ian Affleck and Elliott H. Lieb },
}

@article{YOA1997,
  title = {Nonperturbative Approach to Luttinger's Theorem in One Dimension},
  author = {Yamanaka, Masanori and Oshikawa, Masaki and Affleck, Ian},
  journal = {Phys. Rev. Lett.},
  volume = {79},
  issue = {6},
  pages = {1110--1113},
  numpages = {0},
  year = {1997},
  month = {Aug},
  publisher = {American Physical Society},
  doi = {10.1103/PhysRevLett.79.1110},
  url = {https://link.aps.org/doi/10.1103/PhysRevLett.79.1110}
}

@article {Koma2000,
	author = {Tohru Koma},
	title = {Spectral Gaps of Quantum Hall Systems with Interactions},
	volume = {99},
	pages = {313--381},
	year = {2000},
	URL = {https://doi.org/10.1023/A:1018604925491},
	journal = {J. Stat. Phys.}
}

@inbook{Tasaki2022inbook,
  title={The Lieb-Schultz-Mattis theorem. A topological point of view},
  author={Hal Tasaki},
  booktitle={The Physics and Mathematics of Elliott Lieb},
  volume={II},
  pages={405-446},
  year={2022},
  publisher={EMS Press}
}

@article{KSV1993,
  title = {Theory of polarization of crystalline solids},
  author = {King-Smith, R. D. and Vanderbilt, David},
  journal = {Phys. Rev. B},
  volume = {47},
  issue = {3},
  pages = {1651--1654},
  numpages = {0},
  year = {1993},
  month = {Jan},
  publisher = {American Physical Society},
  doi = {10.1103/PhysRevB.47.1651},
  url = {https://link.aps.org/doi/10.1103/PhysRevB.47.1651}
}

@article{VKS1993,
  title = {Electric polarization as a bulk quantity and its relation to surface charge},
  author = {Vanderbilt, David and King-Smith, R. D.},
  journal = {Phys. Rev. B},
  volume = {48},
  issue = {7},
  pages = {4442--4455},
  numpages = {0},
  year = {1993},
  month = {Aug},
  publisher = {American Physical Society},
  doi = {10.1103/PhysRevB.48.4442},
  url = {https://link.aps.org/doi/10.1103/PhysRevB.48.4442}
}

@Inbook{Resta2007,
author="Resta, Raffaele
and Vanderbilt, David",
title="Theory of Polarization: A Modern Approach",
bookTitle="Physics of Ferroelectrics: A Modern Perspective",
year="2007",
publisher="Springer Berlin Heidelberg",
address="Berlin, Heidelberg",
pages="31--68",
isbn="978-3-540-34591-6",
doi="10.1007/978-3-540-34591-6_2",
url="https://doi.org/10.1007/978-3-540-34591-6_2"
}

@article{Nakagawa2016,
  title = {Flux quench in a system of interacting spinless fermions in one dimension},
  author = {Nakagawa, Yuya O. and Misguich, Gr\'egoire and Oshikawa, Masaki},
  journal = {Phys. Rev. B},
  volume = {93},
  issue = {17},
  pages = {174310},
  numpages = {8},
  year = {2016},
  month = {May},
  publisher = {American Physical Society},
  doi = {10.1103/PhysRevB.93.174310},
  url = {https://link.aps.org/doi/10.1103/PhysRevB.93.174310}
}

@article{LR1972,
  title = {The finite group velocity of quantum spin systems},
  author = {Lieb, E. H. and Robinson, D. W.},
  journal = {Commun. Math. Phys.},
  volume = {28},
  pages = {251--257},
  year = {1972},
  doi = {https://doi.org/10.1007/BF01645779}
}

@misc{NS2010,
title = {Lieb-Robinson Bounds in Quantum Many-Body Physics},
author = {Bruno Nachtergaele and Robert Sims},
note = {arXiv.1004.2086.}
}

@article{BHV2006,
  title = {Lieb-Robinson Bounds and the Generation of Correlations and Topological Quantum Order},
  author = {Bravyi, S. and Hastings, M. B. and Verstraete, F.},
  journal = {Phys. Rev. Lett.},
  volume = {97},
  issue = {5},
  pages = {050401},
  numpages = {4},
  year = {2006},
  month = {Jul},
  publisher = {American Physical Society},
  doi = {10.1103/PhysRevLett.97.050401},
  url = {https://link.aps.org/doi/10.1103/PhysRevLett.97.050401}
}

@article{White1992,
  title = {Density matrix formulation for quantum renormalization groups},
  author = {White, Steven R.},
  journal = {Phys. Rev. Lett.},
  volume = {69},
  issue = {19},
  pages = {2863--2866},
  numpages = {0},
  year = {1992},
  month = {Nov},
  publisher = {American Physical Society},
  doi = {10.1103/PhysRevLett.69.2863},
  url = {https://link.aps.org/doi/10.1103/PhysRevLett.69.2863}
}

@article{DMRG2005,
  title = {The density-matrix renormalization group},
  author = {Schollw\"{o}ck, U.},
  journal = {Rev. Mod. Phys.},
  volume = {77},
  issue = {1},
  pages = {259--315},
  numpages = {0},
  year = {2005},
  month = {Apr},
  publisher = {American Physical Society},
  doi = {10.1103/RevModPhys.77.259},
  url = {https://link.aps.org/doi/10.1103/RevModPhys.77.259}
}

@article{Haegeman2011,
  title = {Time-Dependent Variational Principle for Quantum Lattices},
  author = {Haegeman, Jutho and Cirac, J. Ignacio and Osborne, Tobias J. and Pi\ifmmode \check{z}\else \v{z}\fi{}orn, Iztok and Verschelde, Henri and Verstraete, Frank},
  journal = {Phys. Rev. Lett.},
  volume = {107},
  issue = {7},
  pages = {070601},
  numpages = {5},
  year = {2011},
  month = {Aug},
  publisher = {American Physical Society},
  doi = {10.1103/PhysRevLett.107.070601},
  url = {https://link.aps.org/doi/10.1103/PhysRevLett.107.070601}
}

@article{Haegeman2016,
  title = {Unifying time evolution and optimization with matrix product states},
  author = {Haegeman, Jutho and Lubich, Christian and Oseledets, Ivan and Vandereycken, Bart and Verstraete, Frank},
  journal = {Phys. Rev. B},
  volume = {94},
  issue = {16},
  pages = {165116},
  numpages = {10},
  year = {2016},
  month = {Oct},
  publisher = {American Physical Society},
  doi = {10.1103/PhysRevB.94.165116},
  url = {https://link.aps.org/doi/10.1103/PhysRevB.94.165116}
}

@article{Hauschild2024,
	title = {Tensor network Python (TeNPy) version 1},
	pages = {41},
	author = {Hauschild, Johannes and Unfried, Jakob and Anand, Sajant and Andrews, Bartholomew and Bintz, Marcus and Borla, Umberto and Divic, Stefan and Drescher, Markus and Geiger, Jan and Hefel, Martin and Hémery, Kévin and Kadow, Wilhelm and Kemp, Jack and Kirchner, Nico and Liu, Vincent S. and Moller, Gunnar and Parker, Daniel and Rader, Michael and Romen, Anton and Scalet, Samuel and Schoonderwoerd, Leon and Schulz, Maximilian and Soejima, Tomohiro and Thoma, Philipp and Wu, Yantao and Zechmann, Philip and Zweng, Ludwig and Mong, Roger and Zaletel, Michael P. and Pollmann, Frank},
	journal = {SciPost Phys. Codebases},
	year = {2024},
	publisher = {SciPost},
	doi = {10.21468/SciPostPhysCodeb.41},
	url = {https://scipost.org/10.21468/SciPostPhysCodeb.41}
}

@article{Yokoyama2026,
  title = {Diagnosing energy gap in quantum spin liquids via polarization amplitude},
  author = {Yokoyama, Takayuki and Tada, Yasuhiro},
  journal = {Phys. Rev. B},
  volume = {113},
  issue = {17},
  pages = {174439},
  numpages = {9},
  year = {2026},
  month = {May},
  publisher = {American Physical Society},
  doi = {10.1103/nrbk-mqgz},
  url = {https://link.aps.org/doi/10.1103/nrbk-mqgz}
}

@article{CalabreseCardy2005,
doi = {10.1088/1742-5468/2005/04/P04010},
url = {https://doi.org/10.1088/1742-5468/2005/04/P04010},
year = {2005},
month = {apr},
publisher = {},
volume = {2005},
number = {04},
pages = {P04010},
author = {Calabrese, Pasquale and Cardy, John},
title = {Evolution of entanglement entropy in one-dimensional systems},
journal = {Journal of Statistical Mechanics: Theory and Experiment}
}

@article{CalabreseCardy2007,
doi = {10.1088/1742-5468/2007/10/P10004},
url = {https://doi.org/10.1088/1742-5468/2007/10/P10004},
year = {2007},
month = {oct},
publisher = {},
volume = {2007},
number = {10},
pages = {P10004},
author = {Calabrese, Pasquale and Cardy, John},
title = {Entanglement and correlation functions following a local quench: a conformal field theory
approach},
journal = {Journal of Statistical Mechanics: Theory and Experiment}
}

@article{EislerPeschel2007,
doi = {10.1088/1742-5468/2007/06/P06005},
url = {https://doi.org/10.1088/1742-5468/2007/06/P06005},
year = {2007},
month = {jun},
publisher = {},
volume = {2007},
number = {06},
pages = {P06005},
author = {Eisler, Viktor and Peschel, Ingo},
title = {Evolution of entanglement after a local quench},
journal = {Journal of Statistical Mechanics: Theory and Experiment}
}

@article{Gruber2021,
	title = {Entanglement spreading after local fermionic excitations in the XXZ chain},
	pages = {005},
	author = {Gruber, Matthias and Eisler, Viktor},
	journal = {SciPost Phys.},
	volume = {10},
	year = {2021},
	publisher = {SciPost},
	doi = {10.21468/SciPostPhys.10.1.005},
	url = {https://scipost.org/10.21468/SciPostPhys.10.1.005}
}

@article{Lauchli2008,
doi = {10.1088/1742-5468/2008/05/P05018},
url = {https://doi.org/10.1088/1742-5468/2008/05/P05018},
year = {2008},
month = {may},
publisher = {},
volume = {2008},
number = {05},
pages = {P05018},
author = {Läuchli, Andreas M and Kollath, Corinna},
title = {Spreading of correlations and entanglement after a quench in the one-dimensional
Bose–Hubbard model},
journal = {Journal of Statistical Mechanics: Theory and Experiment}
}

@article{Ramos2024,
  title = {Nonlinear Effects on Charge Fractionalization in Critical Chains},
  author = {Ramos, Fl\'avia B. and Pereira, Rodrigo G. and Eggert, Sebastian and Schneider, Imke},
  journal = {Phys. Rev. Lett.},
  volume = {133},
  issue = {8},
  pages = {086502},
  numpages = {7},
  year = {2024},
  month = {Aug},
  publisher = {American Physical Society},
  doi = {10.1103/PhysRevLett.133.086502},
  url = {https://link.aps.org/doi/10.1103/PhysRevLett.133.086502}
}

@article{Hikihara2008,
  title = {Vector chiral and multipolar orders in the spin-$\frac{1}{2}$ frustrated ferromagnetic chain in magnetic field},
  author = {Hikihara, Toshiya and Kecke, Lars and Momoi, Tsutomu and Furusaki, Akira},
  journal = {Phys. Rev. B},
  volume = {78},
  issue = {14},
  pages = {144404},
  numpages = {19},
  year = {2008},
  month = {Oct},
  publisher = {American Physical Society},
  doi = {10.1103/PhysRevB.78.144404},
  url = {https://link.aps.org/doi/10.1103/PhysRevB.78.144404}
}

@article{Wen1990,
  title = {Chiral Luttinger liquid and the edge excitations in the fractional quantum Hall states},
  author = {Wen, X. G.},
  journal = {Phys. Rev. B},
  volume = {41},
  issue = {18},
  pages = {12838--12844},
  numpages = {0},
  year = {1990},
  month = {Jun},
  publisher = {American Physical Society},
  doi = {10.1103/PhysRevB.41.12838},
  url = {https://link.aps.org/doi/10.1103/PhysRevB.41.12838}
}

@article{GuerreroSuarez2025,
  author  = {Guerrero-Suarez, Ramon and Suresh, Adithya and Maiti, Tanmay and Liang, Shuang and Nakamura, James and Gardner, Geoffrey and Chamon, Claudio and Manfra, Michael},
  title   = {Universal anyon tunnelling in a chiral Luttinger liquid},
  journal = {Nature Physics},
  year    = {2025},
  volume  = {21},
  number  = {11},
  pages   = {1787--1793},
  issn    = {1745-2481},
  doi     = {10.1038/s41567-025-03039-9},
  url     = {https://doi.org/10.1038/s41567-025-03039-9},
}

@article{Langer2011,
  title = {Real-time energy dynamics in spin-$\frac{1}{2}$ Heisenberg chains},
  author = {Langer, Stephan and Heyl, Markus and McCulloch, Ian P. and Heidrich-Meisner, Fabian},
  journal = {Phys. Rev. B},
  volume = {84},
  issue = {20},
  pages = {205115},
  numpages = {14},
  year = {2011},
  month = {Nov},
  publisher = {American Physical Society},
  doi = {10.1103/PhysRevB.84.205115},
  url = {https://link.aps.org/doi/10.1103/PhysRevB.84.205115}
}

@article{Langer2009,
  title = {Real-time study of diffusive and ballistic transport in spin-$\frac{1}{2}$ chains using the adaptive time-dependent density matrix renormalization group method},
  author = {Langer, S. and Heidrich-Meisner, F. and Gemmer, J. and McCulloch, I. P. and Schollw{\"o}ck, U.},
  journal = {Phys. Rev. B},
  volume = {79},
  issue = {21},
  pages = {214409},
  numpages = {11},
  year = {2009},
  month = {Jun},
  publisher = {American Physical Society},
  doi = {10.1103/PhysRevB.79.214409},
  url = {https://link.aps.org/doi/10.1103/PhysRevB.79.214409}
}

@article{Karrasch2014,
  title = {Real-time and real-space spin and energy dynamics in one-dimensional spin-$\frac{1}{2}$ systems induced by local quantum quenches at finite temperatures},
  author = {Karrasch, C. and Moore, J. E. and Heidrich-Meisner, F.},
  journal = {Phys. Rev. B},
  volume = {89},
  issue = {7},
  pages = {075139},
  numpages = {12},
  year = {2014},
  month = {Feb},
  publisher = {American Physical Society},
  doi = {10.1103/PhysRevB.89.075139},
  url = {https://link.aps.org/doi/10.1103/PhysRevB.89.075139}
}

@article{Karrasch2017,
  title = {Proposal for measuring the finite-temperature Drude weight of integrable systems},
  author = {Karrasch, C. and Prosen, T. and Heidrich-Meisner, F.},
  journal = {Phys. Rev. B},
  volume = {95},
  issue = {6},
  pages = {060406(R)},
  numpages = {6},
  year = {2017},
  month = {Feb},
  publisher = {American Physical Society},
  doi = {10.1103/PhysRevB.95.060406},
  url = {https://link.aps.org/doi/10.1103/PhysRevB.95.060406}
}

@article{Al-Hassanieh2013,
  title = {Wave-packet dynamics in the one-dimensional extended Hubbard model},
  author = {Al-Hassanieh, K. A. and Rinc{\'o}n, Juli{\'a}n and Dagotto, E. and Alvarez, G.},
  journal = {Phys. Rev. B},
  volume = {88},
  issue = {4},
  pages = {045107},
  numpages = {6},
  year = {2013},
  month = {Jul},
  publisher = {American Physical Society},
  doi = {10.1103/PhysRevB.88.045107},
  url = {https://link.aps.org/doi/10.1103/PhysRevB.88.045107}
}

@article{Hess2007,
author={Hess, C.},
title={Heat conduction in low-dimensional quantum magnets},
journal={The European Physical Journal Special Topics},
year={2007},
month={Dec},
day={01},
volume={151},
number={1},
pages={73-83},
issn={1951-6401},
doi={10.1140/epjst/e2007-00363-8},
url={https://doi.org/10.1140/epjst/e2007-00363-8}
}

@article{Sologubenko2007,
author={Sologubenko, A. V.
and Lorenz, T.
and Ott, H. R.
and Freimuth, A.},
title={Thermal Conductivity via Magnetic Excitations in Spin-Chain Materials},
journal={Journal of Low Temperature Physics},
year={2007},
month={May},
day={01},
volume={147},
number={3},
pages={387-403},
issn={1573-7357},
doi={10.1007/s10909-007-9317-x},
url={https://doi.org/10.1007/s10909-007-9317-x}
}
\end{document}